\documentclass[prd,superscriptaddress,superscriptaddress,%
twocolumn,showpacs,preprintnumbers,amsmath,amssymb]{revtex4-1}
\usepackage{graphicx,braket}
\usepackage[breaklinks, colorlinks=true, pdfstartview=FitV,%
 linkcolor=red, citecolor=blue, urlcolor=blue]{hyperref}
\usepackage{color}

\def\square{\vcenter{\vbox{\hrule height.4pt
            \hbox{\vrule width.4pt height8pt
            \kern8pt\vrule width.4pt}\hrule height.4pt}}}

\def\sumint{\hbox{$\sum$}\!\!\!\!\!\!\,{\int}}

\newcommand{\MeV}{\,\text{MeV}}
\newcommand{\GeV}{\,\text{GeV}}

\newcommand{\tr}{\text{tr}}
\newcommand{\Nc}{N_{\text{c}}}
\newcommand{\mG}{m_{\text{G}}}
\newcommand{\mGz}{m_{\text{G}0}}
\newcommand{\nB}{n_{\text{B}}}

\newcommand{\beq}{\begin{equation}}
\newcommand{\eeq}{\end{equation}}
\newcommand{\bqa}{\begin{eqnarray}}
\newcommand{\eqa}{\end{eqnarray}}

\newcommand{\p}{\partial}

\begin{document}

\preprint{BI-TP 2013/10}

\title{Stabilizing perturbative Yang-Mills thermodynamics
 with Gribov quantization}

\author{Kenji Fukushima}
\affiliation{Department of Physics, Keio University,
             Kanagawa 223-8522, Japan}

\author{Nan Su}
\affiliation{Faculty of Physics, Bielefeld University,
             33615 Bielefeld, Germany}


\begin{abstract}
We evaluate the thermodynamic quantities of Yang-Mills
theory using the Gribov quantization, which deals with nonperturbative
resummation.  The magnetic scale is automatically incorporated into the
framework and we find it efficient to stabilize the perturbative
expansion of the free energy.  In the temperature range
$T=T_c \!\sim\! 2\,T_c$ the major uncertainty in our results comes from
the nonperturbative running coupling that is adopted from the lattice
simulation, while the convergence above $2\,T_c$ is impressively
robust.  We also present the corresponding interaction measure
(i.e., trace anomaly) up to close to $T_c$.
\end{abstract}

\maketitle

\section{Introduction}
Stimulated by the exciting developments from the Relativistic Heavy
Ion Collider (RHIC) and Large Hadron Collider (LHC) experiments,
the thermodynamics of the quark-gluon plasma---especially the pressure $p$
or the free energy density $f=-p$---is of crucial interest.  Due to
asymptotic freedom~\cite{asymptotic-freedom}, perturbation theory was
expected to work for QCD thermodynamics at high enough
temperatures.  The weak-coupling expansion of the QCD free energy has been
accomplished up to $g^6\log g$, which is the highest order in the
perturbative approach, with $g$ being the running coupling
(see Ref.~\cite{reviews} for reviews).  Unexpectedly, the resulting
weak-coupling series shows poor convergence in the intermediate-temperature 
regime, i.e., $T=2\, T_c \!\sim\! 4\, T_c$, with
$T_c\sim160\MeV$ being the pseudocritical
temperature for the QCD phase transition (or $T_c\sim270\MeV$ for pure Yang-Mills theory, 
abbreviated as YM hereafter), which is probed in the RHIC and
LHC experiments.  Resummation must be carried out to incorporate the
contributions from the electric scale
$gT$~\cite{bir,eqcd,htlpt,htlpt-ym}.

The nonconvergence of QCD thermodynamics is attributed to the
IR sector of QCD, which is governed by soft-scale
gluons~\cite{reviews}.  Therefore the pure Yang-Mills theory
provides the simplest test bench to access the IR problem.  While the
resummed perturbation theory in the electric sector resulted in
improved convergence of the YM free energy down to $T\sim3\, T_c$,
significant deviations from the lattice data were observed by
lowering $T$ further toward $T_c$, which could be taken as a signal of the
onset of nonperturbative effects.

It was discovered in the early 1980s that due to the absence of an IR
cutoff by screening in the magnetic sector, the perturbative expansion of
finite-$T$ YM theory breaks down at a fundamental level at the
magnetic scale $g^2T$.  This IR catastrophe is the so-called Linde
problem~\cite{linde-problem}.  As a result, perturbation theory works
up to a certain order only depending on the quantity in consideration.
For the free energy IR divergences from the magnetic scale start
entering at four loops, corresponding to $g^6$ order, which makes the
highest accessible order $g^6\log g$, as mentioned before.  In the RHIC
and LHC temperature regime the running coupling is $g\sim{\cal O}(1)$,
and therefore the missing contributions from the nonperturbative magnetic
scale should be as significant as the ones from the perturbative
electric scale.  This fact provides us with an explanation for the
aforementioned deviations in the free energy of resummed perturbation
theory from the lattice data.  There have been various attempts to
incorporate the magnetic contributions, but the nonperturbative
nature is inherent in the confining properties of dimensionally reduced YM
theory at high temperature~\cite{mag_mass}.  We must thus augment the
resummed perturbation theory with a confinement mechanism even when we
deal with a deconfined state of matter.

The key ingredients in constructing the YM thermodynamics are the
correlation functions of the gluons and the ghosts.  Gauge fixing is
conveniently done through the Faddeev-Popov (FP) procedure~\cite{fp},
and the bare gluon and ghost propagators show an IR pole proportional
to $1/p^2$, with $p=(p_0,\boldsymbol{p})$ being the Minkowskian
four-momentum.  The corresponding on-shell dispersion relation is
$p_0=|\boldsymbol{p}|$, which is modified in the full propagators
including interaction effects.

In order to have a better understanding in the intermediate-temperature 
regime, the YM correlation functions must be improved
nonperturbatively, including the magnetic sector.  The last decade has
witnessed tremendous developments of high-precision lattice
measurements of the YM correlation functions at both zero and finite
temperatures~\cite{Maas:2011se}.  There has also been considerable
progress in the study of YM correlation functions by
functional methods in, e.g., Ref.~\cite{von Smekal:1997is}, and the results
are in good agreement with the lattice simulation
(see Ref.~\cite{functional} and references therein).  The obtained gluon
and ghost propagators show an IR suppression and enhancement,
respectively, as compared to the FP case.  These results heuristically
encompass desirable features of confinement:  the IR-suppressed gluon
propagator indicates gluon confinement at large distance, and the
IR-enhanced ghost is responsible for confinement.  It has also been
demonstrated that the balance in the IR sector is responsible for the
deconfinement phase transition~\cite{Braun:2007bx,Fukushima:2012qa}.

Gribov pointed out more than three decades ago that there still
remains a residual gauge ambiguity in the IR sector, i.e., the
so-called Gribov copy problem~\cite{Gribov:1977wm}.  If the functional
integration is dominated by the contributions near the horizon of the
fundamental modular region, the ghost propagator is naturally
enhanced in the IR momenta, and the resulting gluons are suppressed in
turn.  This is in line with the results from the lattice and
functional methods.  The corresponding on-shell dispersion relation
reads $p_0=\sqrt{\boldsymbol{p}^2+\mG^4/\boldsymbol{p}^2}$, with $\mG$
being the Gribov mass parameter which is solved by the condition of
the horizon dominance.  It was shown by Zwanziger in a
phenomenological way that a free gas of Gribov quasiparticles
qualitatively captures the nonperturbative features of the lattice
equation of state~\cite{Zwanziger:2004np}.  Subsequently, it was found
that the Gribov mass parameter is correctly proportional to the
magnetic scale, i.e., $\mG \sim g^2T$, in the limit
$T\rightarrow\infty$~\cite{Zwanziger:2006sc}.  This finding is
promising enough and a confining mechanism can indeed resolve the Linde
problem by providing a nonperturbatively generated IR cutoff.
Many of the works on the Gribov copy problem are dedicated to the
vacuum~(see Ref.~\cite{Vandersickel:2012tz} for a review).  Systematic
attempts to extend the theory of confinement to the finite-$T$
problems should deserve further investigations along the same line as
a first try~\cite{Zwanziger:2006sc}.  In this work we will pursue this
possibility to make the perturbative evaluation of YM free energy
stabilized by the nonperturbative mass scale emerging from the Gribov
gauge-fixing procedure in which features of confinement are
implemented, which we simply call the Gribov quantization.

\section{Formalism}
In the Gribov quantization~\cite{Gribov:1977wm}, the YM partition
function in Euclidean space reads
\begin{equation}
 Z = \int_\Omega \mathcal{D}A(x)\, V(\Omega)\,
 \delta( \p \!\cdot\! A )\, \det[ -\p \!\cdot\! D(A) ]\; e^{-S_{\rm YM}} \;,
\label{eq:Z_YM}
\end{equation}
in which the functional integration should be carried out within the
Gribov region defined as
\begin{equation}
 \Omega \equiv \{  A:  \p \cdot A = 0 ,  -\p \cdot D(A) \geq 0 \} \;,
\end{equation}
whereby the Landau gauge is chosen.  The restriction of the
integration to the Gribov region is realized by inserting a function
$V(\Omega)$ into the partition function~\eqref{eq:Z_YM}, where
\begin{equation}
 V(\Omega) = \theta [ 1 - \sigma(0) ]
  = \int_{-i\infty+\epsilon}^{+i\infty+\epsilon}
  \frac{d\beta}{2\pi i \beta} \, e^{\beta[ 1 - \sigma(0) ]}
\end{equation}
represents the no-pole condition.  Here, $1 - \sigma(P)$ is the
inverse of the ghost dressing function $Z_G(P)$ [see
  Eq.~(\ref{eq:ghost})].  The integration variable $\beta$ is
identified as the Gribov mass parameter $\mG$ after a redefinition
(see Ref.~\cite{Vandersickel:2012tz} for technical details of the Gribov
quantization).

\subsection{Gap equation}
The Gribov mass parameter $\mG$ is a Lagrange multiplier that is
determined by the variational principle, leading to the following
\textit{gap equation}:
\begin{equation}
 \sumint_{\!\!\!P}\; {1 \over P^4 + \mG^4}
  = {d \over (d-1) \Nc g^2 } \;,
\label{eq:gap}  
\end{equation}
where the Euclidean four-momentum reads $P=(\boldsymbol{p},p_4)$, with
$p_4=2n\pi T$, and $N_c$ is the number of colors.  Our calculation is
carried out in dimensional regularization with the $\overline{\rm MS}$
renormalization scheme, in which the sum-integral is defined as
\begin{equation}
\sumint_{P} \equiv
  \left(e^{\gamma_E}\mu^2 \over 4\pi\right)^\epsilon \;
  T\sum_{p_4=2n\pi T}\:\int {d^{d-1}p \over(2 \pi)^{d-1}} \;,
\end{equation} 
with $d=4-2\epsilon$ being the spacetime dimensions.

After carrying out the sum-integral and subtracting the UV divergence
(see Ref.~\cite{Gracey:2005cx} for details) the gap equation becomes
\bqa\nonumber
 1 \,&=&\, \frac{3\Nc g^2}{64\pi^2} \biggl[ \frac{5}{6}
  - \ln\biggl(\frac{\mG^2}{\mu^2}\biggr) \\
 &&
 +\,{4 \over i\mG^2} \int_0^\infty dp\,p^2 
\left( {n_B(\omega_{-}) \over \omega_{-}}
 - {n_B(\omega_{+}) \over \omega_{+}} \right)
  \biggr ] \;,
\label{eq:gap1}
\eqa
where $\omega_\pm=\sqrt{p^2\pm i\mG^2}$ and
$\nB(x)\equiv(e^{x/T}-1)^{-1}$ is the Bose-Einstein distribution
function.  Analytical solutions are available in the limiting cases
as
\begin{equation}
 \mG = \begin{cases}
  \displaystyle \mu\, \exp\biggl( \frac{5}{12}
  - \frac{32\pi^2}{3\Nc g^2} \biggr) & (T\to 0)\;, \\[1em]
  \displaystyle \frac{d-1}{d}\cdot\frac{\Nc}{4\sqrt{2}\pi} g^2 T &
  (T\to \infty)\;,
  \end{cases}
 \label{eq:mG}
\end{equation}
in which it is evident that the magnetic scale emerges at high
temperature.  In what follows we will solve Eq.~\eqref{eq:gap1}
numerically to derive $\mG$ as a function of $T$ for a given
renormalization scale $\mu$ (which will be fixed later).

\subsection{Gluon propagator}
Gribov's gluon propagator in the Landau gauge reads
\begin{equation}
 D_A (P) = \delta^{ab} {P^2 \over P^4 + \mG^4} \biggl(
  \delta^{\mu\nu} - {P^\mu P^\nu \over P^2} \biggr) \;,
\label{eq:gluon_prop}
\end{equation}
which is regular as $P\to0$, indicating the IR suppression of confined
gluons~\cite{Gribov:1977wm}.  To proceed to any calculation involving
$D_A(P)$, we need to specify the running coupling as a function of
$T$.  Since we are interested in the IR regime, the perturbative
running is not quite appropriate and we have to use the full
nonperturbative running.  Instead of solving the nonperturbative
resummation, we adopt the lattice results at finite
$T$~\cite{Kaczmarek:2004gv}.  In Ref.~\cite{Kaczmarek:2004gv} the running
coupling is extracted from the large-distance (IR) and short-distance
(UV) behaviors of the heavy-quark free energy, leading to
$\alpha_{\text{IR}}$ and $\alpha_{\text{UV}}$, respectively.

\begin{figure}
 \includegraphics[width=\columnwidth]{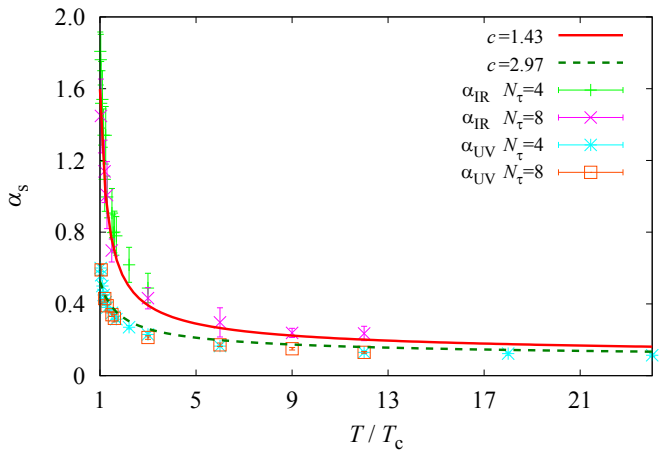}
 \caption{Lattice data of IR (upper) and UV (lower) running couplings
   at finite $T$ from Ref.~\cite{Kaczmarek:2004gv} and the fit results with
   a choice of $c$ in Eq.~\eqref{eq:running}.}
 \label{fig:running}
\end{figure}

Interestingly, the lattice-measured coupling can be nicely fitted by a
one-parameter form as
\begin{equation}
 \alpha_s(T / T_c) \equiv \frac{g^2(T / T_c)}{4\pi}
 = \frac{6\pi}{11\Nc \ln{[c(T / T_c)]}} \;,
 \label{eq:running}
\end{equation}
with $c=1.43$ for the IR case and $c=2.97$ for the UV case, as shown
together with the lattice data in Fig.~\ref{fig:running}.  In the
intermediate-temperature regime, the running coupling has a
substantial scheme dependence and we therefore use the IR and UV
values to estimate theoretical uncertainties.

\subsection{Ghost propagator}
The no-pole condition (i.e., to not go across the Gribov horizon) requires
that the IR limit of the ghost propagator should be enhanced,
indicating the proximity to the Gribov horizon~\cite{Gribov:1977wm}.
The ghost propagator in the Landau gauge thus reads
\begin{equation}
 D_c(P) = \delta^{ab} {1 \over 1-\sigma(P)}\cdot {1 \over P^2} \;,
\end{equation}
where the ghost dressing function is $Z_G\equiv [1-\sigma(P)]^{-1}$
and $\sigma(P)$ with gluon ladder diagrams taken into account turns
out to be
\begin{equation}
 \sigma(P) \equiv \Nc g^2 {P^\mu P^\nu \over P^2} \sumint_{\!\!\!Q}
  {1 \over Q^4 \!+\! \mG^4}{Q^2 \over (Q \!-\! P)^2}
  \biggl( \delta^{\mu\nu} - {Q^\mu Q^\nu \over Q^2} \biggr) \;.
\label{eq:ghost} 
\end{equation}
Using the gap equation~\eqref{eq:gap} one can confirm $1-\sigma(0)=0$,
indicating the IR enhancement of the ghosts.  We note that recent
lattice data favor the so-called decoupling solution for which the
ghost dressing function stays finite at $P\to0$, which can be
incorporated into a refined Gribov-Zwanziger
approach~\cite{Dudal:2008sp}.  In this study, however, we keep using
the original form (shown above) for simplicity. Different behaviors in the
deep IR region hardly affect bulk thermodynamics. In order to verify
this, we have inserted a mass parameter to modify the deep-IR behavior
of the propagators in favor of the decoupling solution, and have found
that the resulting pressure decreases at most less than $10\%$ at
$T=T_c$, and only a few percent at $T>T_c$.

\begin{figure}
 \includegraphics[width=\columnwidth]{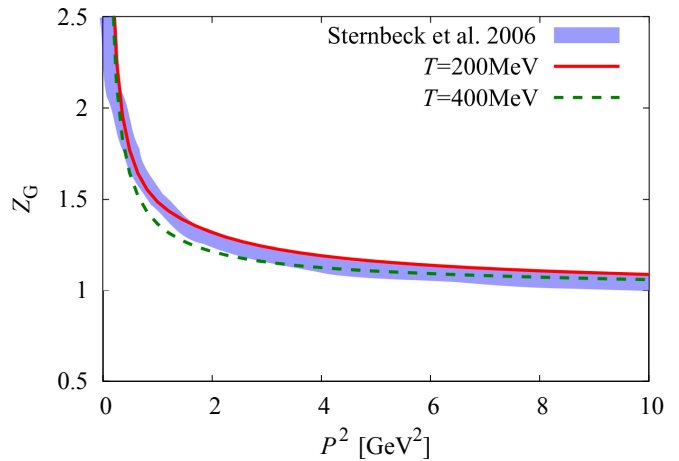}
 \caption{Ghost dressing functions at $T=200\MeV$ (solid curve) and
   $T=400\MeV$ (dashed curve) as compared to the lattice data from
   Ref.~\cite{Sternbeck:2006cg}.}
 \label{fig:zg}
\end{figure}

At $T=0$ this integration can be performed with dimensional
regularization, yielding
\begin{align}
 & 1-\sigma(P) = {\Nc g_0^2 \over 128\pi^2} \biggl[ -5
 + \biggl(3-{\mGz^4 \over P^4}\biggr)\ln\biggl( 1 + {P^4 \over \mGz^4}
   \biggr) \notag \\
 &\qquad +{\pi P^2 \over \mGz^2} + 2\biggl(3-{P^4 \over \mGz^4}\biggr)
  {\mGz^2 \over P^2}\arctan{P^2 \over \mGz^2} \biggr] \;,
\label{eq:ghost_vac}
\end{align}
where $\mGz=\mG(T=0)$, and we find that a choice of $\mu=1.69\GeV$ [in
view of Eq.~\eqref{eq:mG} with $g=3.13$ adjusted by hand] can
reproduce the $T=0$ lattice data of
$Z_G$~\cite{Sternbeck:2006cg,lat-ghost}.  Amazingly, with the same
$\mu$ fixed at $T=0$, our finite-$T$ numerical results from
Eq.~\eqref{eq:ghost} [with $g(T)$ from Eq.~\eqref{eq:running}] are
rather insensitive to $T$, which is perfectly consistent with the
lattice simulation.  We note that in the direct evaluation of
Eq.~\eqref{eq:ghost} we imposed a three-momentum cutoff
$\Lambda=1.25\mu$ so that the $T=0$ numerical results are matched
with Eq.~\eqref{eq:ghost_vac}.  Figure~\ref{fig:zg} shows the
numerical results at $T=200\MeV$ and $T=400\MeV$ when $g(T)$ runs with
$c=2.97$.  If we use $c=1.43$, for the $T=200\MeV$ case, $cT$ is too
close to $T_c$.  As long as $T$ is greater than $T_c$, however, the
behavior of $Z_G$ is robust and not contaminated by the uncertainty in
$g(T)$.

Now that we have confirmed that the resummed ghost dressing function
in our approach is almost $T$ independent, as observed in the lattice
simulation, we can safely use Eq.~\eqref{eq:ghost_vac} for the
finite-$T$ calculation too.

\section{Quasiparticle Approximation}
With the gluon and ghost propagators prepared, we can calculate the
free energy.  In the two-particle irreducible (2PI) formalism the
effective action can be expressed in terms of the full propagator $G$
and full self-energy $\Pi$ as
\begin{equation}
 \Gamma = \frac{1}{2}\tr\ln G^{-1} - \frac{1}{2}\tr \, \Pi G
 + \Gamma_2[G] \;,
\label{eq:2PI}
\end{equation}
where $\Gamma_2[G]$ represents the sum of all 2PI diagrams.  In a
general quasiparticle approximation we keep only the first term,
which should be a reasonable estimate;  the variational principle to
derive $G$, i.e., $\delta\Gamma/\delta G=0$, can be formulated
to \textit{minimize} the sensitivity to higher-order corrections,
which is often called optimized perturbation
theory~\cite{Chiku:1998kd}.  This means that the first term in
Eq.~\eqref{eq:2PI} must be dominant as long as $G$ is
self-consistently determined, or equivalently, if we use the full
$G$ that should coincide with the self-consistently determined $G$,
in principle.

\subsection{Gluon contribution}
The gluon contribution to the free energy reads
\beq
 {1 \over 2}\tr\ln D_A^{-1} = {\Nc^2-1 \over 2} \sumint_{\!\!P}
 \biggl[ 3\ln \frac{P^4 + m_G^4}{P^2} + \ln P^2 \biggr] \;,
\eeq
which can be evaluated straightforwardly with dimensional
regularization, giving
\begin{align}
 & \frac{1}{2}\tr\ln D_A^{-1} = (\Nc^2\!-\!1)\biggl\{
  \frac{\pi^2 T^4}{45} + \frac{3\mG^4}{32\pi^2} \biggl(
  \frac{3}{2}-\ln\frac{\mG^2}{\mu^2} \biggr) \notag\\
 &\qquad\qquad +\frac{3T}{2\pi^2}\sum_\pm \int_0^\infty dp\, p^2 \ln(
  1-e^{-\omega_\pm/T}) \biggr\} \;.
\label{eq:gluon_loop}  
\end{align}
We can then calculate the gluon contribution numerically using
  the value of $\mu$ fixed by the ghost propagator and $\mG$ as a
  numerical solution of the gap equation~\eqref{eq:gap1}.

\subsection{Ghost contribution}
The ghost contribution to the free energy reads
\begin{align}
 & -\tr\ln D_c^{-1} = -(\Nc^2-1) \sumint_{\!\!P} \ln \Bigl\{
  P^2 \bigl[\,1-\sigma(P)\,\bigr] \Bigr\} \notag\\
 &= (\Nc^2-1) \biggl\{ {\pi^2T^4 \over 45}
  - \sumint_P \ln\bigl[\, 1-\sigma(P)\,\bigr] \biggr\} \;,
\label{eq:ghost_loop}
\end{align}
with Eq.~\eqref{eq:ghost_vac} substituted for $1-\sigma(P)$, which is
justified numerically.  It is technically daunting to do the
sum-integral of Eq.~\eqref{eq:ghost_loop} analytically, and we will
resort to the numerical calculation, in which the UV divergence is
subtracted at a certain $T$ sufficiently below $T_c$.

\begin{figure}
 \includegraphics[width=0.86\columnwidth]{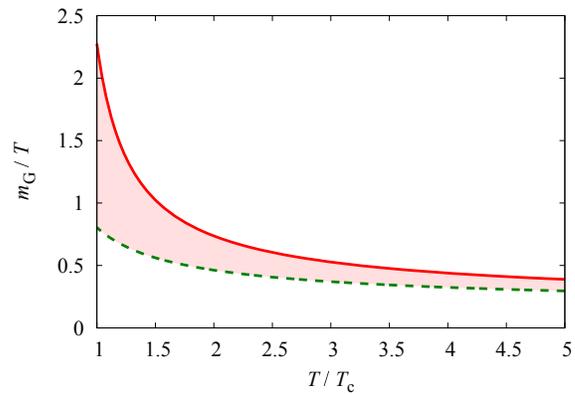}
 \caption{Gribov mass parameter as a function of $T$.  The band shows
   the uncertainty arising from different schemes of lattice-measured
   $\alpha_s(T)$ (see Fig.~\ref{fig:running}).  The upper (solid) and
   lower (dashed) bounds correspond to the IR and UV couplings,
   respectively.}
 \label{fig:mg_t}
\end{figure}

\section{Results and Discussions}
With the renormalization scale $\mu$ fixed as $\mu=1.69\GeV$, we can
numerically solve the gap equation~\eqref{eq:gap1} to obtain $\mG$ as
a function of $T$, which we show in Fig.~\ref{fig:mg_t}.

\begin{figure}
 \includegraphics[width=\columnwidth]{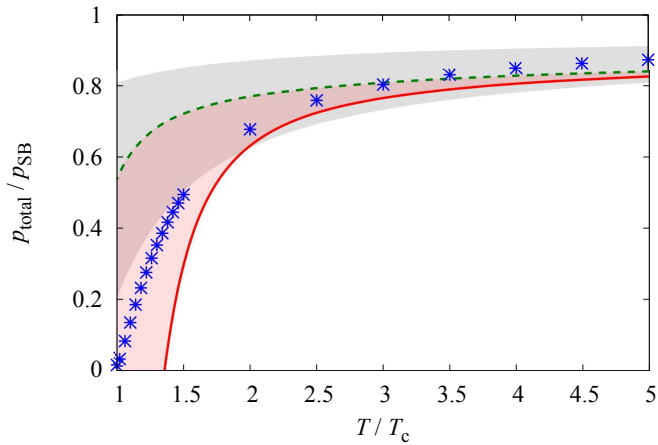}
 \caption{Pressure (free energy): The red band represents our
   numerical results with the solid and dashed curves
   corresponding to those in Figs.~\ref{fig:running} and
   \ref{fig:mg_t}.  The blue dots represent lattice data from
   Ref.~\cite{Borsanyi:2012ve}, and the gray band represents three-loop
   HTLpt results from Ref.~\cite{htlpt-ym}.}
 \label{fig:total}
\end{figure}

Solving the gluon contribution~\eqref{eq:gluon_loop} and the ghost
contribution~\eqref{eq:ghost_loop} numerically with $\mG$ as
determined in Fig.~\ref{fig:mg_t}, we obtain the pressure or the free
energy of Gribov quasiparticles scaled by the massless
Stefan-Boltzmann limit, $-(\Nc^2-1)\pi^2T^4 / 45$ for $\Nc=3$, as shown
in Fig.~\ref{fig:total}.  In technical practice, the ghost
contribution is naively UV divergent and needs regularization and
subtraction.  We regularized the integration by inserting a smooth
function, $(1-\tanh((P^2-P_0^2)/\Delta P^2))/2$, where we chose
$P_0=100\GeV$ and $\Delta P^2/P_0^2=0.4$.  Because we numerically
process the Matsubara sum in a straightforward way, it is necessary to
use a sufficiently smooth regularization in order to avoid unphysical
cutoff artifacts.  Then we extracted the matter part of the ghost
contribution by subtracting the pressure at a temperature far below
$T_c$ (which is $50\MeV$ in this work).  Of course the final results
are insensitive to this choice of the subtraction point, which we have
explicitly checked.

To draw Fig.~\ref{fig:total}, the SU(3) YM lattice data (blue
dots) are taken from Ref.~\cite{Borsanyi:2012ve}.  The red band shows the
uncertainty arising from the lattice running couplings, with the lower
(solid) and upper (dashed) bounds corresponding to the IR and the
UV couplings, respectively, as we have seen in
Fig.~\ref{fig:running}.

It is worth noting that our calculations are done in a way consistent
with the lattice-measured $\alpha_s(T)$, while the perturbative
running coupling with an assumed renormalization scale used in
conventional resummed perturbation theory is not compatible with the
lattice $\alpha_s(T)$ even at very high $T$ (see discussions in
Ref.~\cite{Kaczmarek:2004gv} for more details).  Due to the ambiguity in
defining $\alpha_s(T)$ on the lattice, we see a rather wide band at
$T\lesssim2\,T_c$.  However, the uncertainty in the resulting free
energy gets suppressed significantly at $T\gtrsim2.5\,T_c$, which is
highly nontrivial, especially considering the fact that there is still
a big variation between the IR and UV lattice couplings at such
temperatures.

Comparing to the most recent estimate from three-loop
hard-thermal-loop perturbation theory (HTLpt)~\cite{htlpt-ym}, as shown
by the gray band in Fig.~\ref{fig:total}, the uncertainty in our free
energy is about $35\%$ that of the HTLpt at $2.5\, T_c$, and about
$15\%$ at $5\, T_c$.  We would like to point out, furthermore, that
our free energy is consistent with the lattice data in the whole
displayed temperature range.  In particular the lattice data lie in
our band even below $3.5\,T_c$, where HTLpt suffers from poor
convergence, and therefore a sharp rising of the free energy after the
deconfinement phase transition---which indicates a nonperturbative
release of new degrees of freedom---is realized somehow even without
the inclusion of the Polyakov loop.  The free energies from the
resummation schemes of Refs.~\cite{bir,eqcd} have relatively smaller
uncertainties than HTLpt, but they are not successful enough to
recover the sharp rising behavior below $2\,T_c$, where we consider
that the resummation in the magnetic sector is crucial.

\begin{figure}
 \includegraphics[width=\columnwidth]{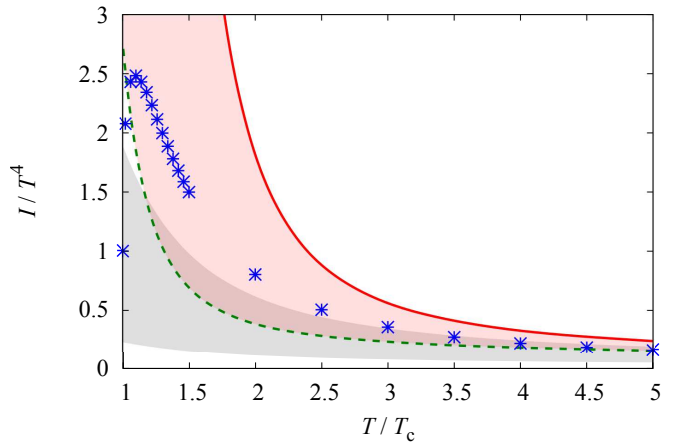}
 \caption{Interaction measure (trace anomaly):  The red band
   represents our numerical results with the solid and dashed
   curves corresponding to those in Figs.~\ref{fig:running},
   \ref{fig:mg_t}, and \ref{fig:total}.  The blue dots represent
   lattice data from Ref.~\cite{Borsanyi:2012ve}, and the gray band
   represents three-loop HTLpt results from Ref.~\cite{htlpt-ym}.}
 \label{fig:trace}
\end{figure}

All the other thermodynamic quantities can be derived from the free
energy and in Fig.~\ref{fig:trace} we show the $T^4$-scaled
interaction measure $I$ (which is sometimes called the trace anomaly),
defined as $I=\epsilon-3p=T^5{d \over dT}(p / T^4)$, where $\epsilon$
is the energy density given by $\epsilon=T dp/dT-p$.  The interaction
measure $I$ vanishes when the theory preserves scale invariance, and
therefore it is an important measure for the breaking of scale invariance
by interaction effects and the quantum transmutation of the mass scale.
Besides, because $I$ involves a derivative with respect to $T$,
the interaction measure is obviously more informative than the
pressure $p$.  In other words, the consistency seen in the pressure---as
in Fig.~\ref{fig:total}---does not guarantee quantitative agreement in
view of $I$.

We present our result in Fig.~\ref{fig:trace}, which shows similar
behavior as the pressure in Fig.~\ref{fig:total}:  the uncertainty
from the lattice $\alpha_s(T)$ gets highly suppressed for increasing
values of $T$ and the lattice data lie in our band for almost the entire regime
shown in the plot.  It is clear that $\mG$ has a sizable effect on the
interaction measure at low $T$ ($\lesssim2\,T_c$) compared to the
three-loop HTLpt result, which indicates the onset of the
nonperturbative magnetic scale.  Due to the absence of the effects of the
Polyakov loop, the peak in the interaction measure---which indicates the onset 
(for decreasing $T$) of phase-transition physics---is not manifest
in our result.

In closing, we would like to stress that the resummation in the
magnetic sector is by itself incorporated through the Gribov quantization,
though the quasiparticle approximation may look like too simple an 
approach.  In fact, however, the construction of $\sigma(P)$ explicitly
involves the higher-order diagrams, and solving the gap equation is
nothing but the variational calculation to carry out the
nonperturbative resummation effectively.  In this sense, it is
neither an accident nor magic that our results are consistent
with the lattice data, but rather a natural consequence from the proper
reformation of the magnetic contributions.

\section{Conclusions and Outlook}
In this paper we have explored a novel and systematic evaluation of
the Yang-Mills free energy using the Gribov quantization.  The results,
thanks to the improvement at the magnetic scale through the
self-consistent solution for $\mG$, show robust and stable behavior
consistent with the lattice data, which is the first self-consistent
calculation that significantly surpasses the first attempt
in Ref.~\cite{Zwanziger:2006sc}.  We have evaluated the finite-$T$ ghost
propagator from the Gribov quantization and shown its
\textit{insensitivity} to $T$, which to our knowledge is the very
first semianalytic result that is in surprising agreement with the
latest lattice observation as seen
in Refs.~\cite{Sternbeck:2006cg,lat-ghost}.  All these results evidently
manifest the profound importance of the nonperturbative gauge fixing even
in the perturbative evaluation of thermodynamics at high temperature.

There are intriguing future extensions.  First, the explicit
calculation of the magnetic screening mass and the spatial string
tension would provide us with a deeper insight into the Linde problem.
Second, the Polyakov-loop coupling would enable us to investigate the
first-order deconfinement transition~\cite{Fukushima:2012qa}.  Third,
the running coupling from the functional approaches 
are qualitatively in line with the lattice result and
would help in reducing the uncertainty band~\cite{Braun:2005uj,jan}.  Fourth, it
would be desirable to systematize the perturbation theory using the
local Gribov-Zwanziger action~\cite{Zwanziger:1989mf}.  Last but not
least, the application of the Gribov quantization would have a
profound impact on more general QCD problems, e.g., the magnetic
properties of QCD matter~\cite{xqcd}, transverse dynamics of real-time
evolution~\cite{Dumitru:2013koh}, and so on.

\begin{acknowledgments}
We are grateful for stimulating discussions with L.~Fister, T.~Kojo,
A.~Maas, J.~M.~Pawlowski, Y.~Schr\"oder, L.~von~Smekal, M.~Strickland,
and K.~Tywoniuk.  We thank O.~Kaczmarek for providing us with data of
\cite{Kaczmarek:2004gv}.  K.~F.\ was supported by JSPS KAKENHI Grant
\# 24740169.  N.~S.\ acknowledges support from the Alexander von
Humboldt Foundation.
\end{acknowledgments}


\end{document}